\documentclass[copyright,creativecommons]{eptcs}

\usepackage {bsymb,b2latex}
\usepackage{epsfig}
\usepackage{graphicx}

\title{Self-Recovering Sensor-Actor Networks}
\author{Maryam Kamali, Linas Laibinis, Luigia Petre, Kaisa Sere
\institute{IT-Department, \AA bo Akademi University, Turku, Finland}
\email{\{maryam.kamali,linas.laibinis,luigia.petre,kaisa.sere\}@abo.fi}}
\def\titlerunning{Self-Recovering Sensor-Actor Networks}
\def\authorrunning{M. Kamali, L. Laibinis, L. Petre \& K. Sere}

\begin{document}
\maketitle

\begin{abstract}
Wireless sensor-actor networks are a recent development of wireless networks where both ordinary sensor nodes and more sophisticated and powerful nodes, called \emph{actors}, are present. In this paper we formalize a recently introduced algorithm that recovers failed actor communication links via the existing sensor infrastructure. We prove via refinement that the recovery is terminating in a finite number of steps and is distributed, thus self-performed by the actors. Most importantly, we prove that the recovery can be done at different levels, via different types of links, such as direct actor links or indirect links between the actors, in the latter case reusing the wireless infrastructure of sensors. This leads to identifying \emph{coordination classes} e.g., for delegating the most security sensitive coordination to the direct actor-actor coordination links, the least real-time constrained coordination to indirect links, and the safety critical coordination to both direct actor links and indirect sensor paths between actors. Our formalization is done using the theorem prover in the RODIN platform.\\

\end{abstract}

\section{Introduction}
The separation of computation and control stands at the basis of the software architecture discipline. The control of the \emph{computing} entities as well as the coordination of the \emph{controlling} entities are well illustrated by Wireless Sensor-Actor Networks (WSANs), a rather new generation of sensor networks~\cite{1}. A WSAN is made of two kinds of nodes: \emph{sensors} (the `computing' entities) and \emph{actors} (the controlling entities), with the density of sensor nodes much bigger than that of actor nodes. The sensors detect the events that occur in the field, gather them and transmit the collected data to the actors. The actors react to the events in the environment based on the received information. The sensor nodes are low-cost, low-power devices equipped with limited communication capabilities, while the actor nodes are usually mobile, more sophisticated and powerful devices compared to the sensor nodes.

\begin{figure}
 \label{Figure1}
 \centering
 \includegraphics[scale=0.6]{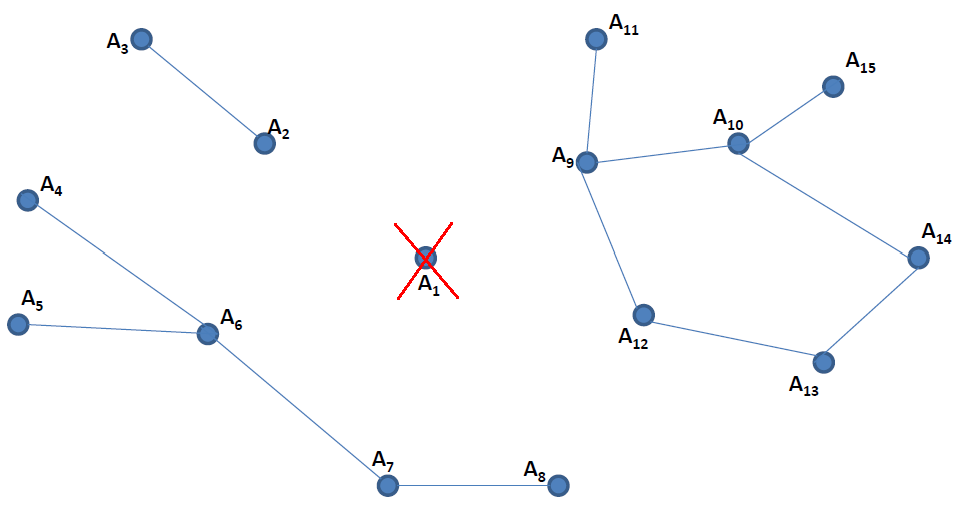}
 \caption{Three partitions created by the failed actor $A_1$}
 \end{figure}

A central WSAN requirement is that of node \emph{coordination}. As there is no centralized control in a WSAN, sensors and actors need to coordinate with each other in order to collect information and take decisions on the next actions~\cite{1}.
There are three main types of WSAN coordination~\cite{2}: sensor-sensor, sensor-actor and actor-actor coordination, the latter being considered in this paper.
The actor-actor coordination is concerned with the actor decisions and the division of tasks among different actors. To achieve the actor-actor coordination in WSANs, actors need reliable connection links for communicating with each other, which are established upon initializing the WSAN. However, WSANs are dynamic networks where the network topology continuously changes. The changes occur when new links or nodes are added and when new links or nodes are removed due to failures, typically generated by hardware crashes, lack of energy, malfunctions, etc. Thus, actor nodes can fail during the operation of the network. As a result, a WSAN may transform into several, disconnected WSAN sub-networks. This separation is called \emph{network partitioning} and is illustrated in Fig.~\ref{Figure1}, where the actor nodes $A_1-A_{15}$ are shown to produce a network partitioning if actor node $A_1$ fails.

Another central WSAN requirement is that of embedding \emph{real-time aspects}. Thus, depending on the application, it might be essential to respond to sensor inputs within predefined time limits, e.g., in critical applications such as forest fire detection. Due to the real-time requirements of WSANs, the failure of an actor node should not impact the whole actor network for too long. The problem of actor failing in the actor-actor coordination has been already addressed. For instance, the physical movement of actor nodes towards each other is proposed in~\cite{3,4} to re-establish their connectivity. However, during this movement, nodes in different partitions that have been created by the actor failure cannot coordinate. To shorten the time of recovery, Kamali et al~\cite{5} have previously proposed an algorithm for establishing new routes between non-failed actors via sensor nodes. This algorithm allows to quickly reconnect the separated partitions, \textit{before} moving actor nodes as proposed in~\cite{3,4}. In this paper, we further study this recovery mechanism that alleviates the actor coordination failure.

There are several properties that are desirable to verify for this algorithm. First, we need to show that there is \emph{always} a path via sensor nodes that \emph{can be} established by the partitioned actor nodes. Second, it is desirable to guarantee that this path is the shortest, in order not to overload the power-limited sensor nodes. Third, to shorten the time of recovery as much as possible, it is desirable to establish the connection as soon as possible. In this paper we demonstrate the first property of the algorithm. 

The contribution of this paper is threefold. First, we formalize the algorithm introduced recently for self-recovering actor coordination, using a theorem prover tool. This allows us to better understand the functioning of the algorithm. Second, we prove the termination of the algorithm, namely we prove that the recovery of a link ends in a  finite number of steps. This proves that the recovery path \emph{can be} established between non-failed actors. An important aspect of the recovery method is that the indirect links between actors (via sensors or not) are built in a \emph{distributed} manner, thus ensuring the \emph{self}-recovering of the network. Since sensors are so numerous in a WSAN, practically covering the area they sense, we thus prove the first desirable property of the algorithm, i.e., that there is \emph{always} a path via sensor nodes that \emph{can be} established by the partitioned actor nodes. Third, we prove that the recovery can be done at different levels, via different types of links, such as direct actor links or indirect links between the actors, in the latter case also reusing the WSAN infrastructure of sensors. In fact, the novelty of our contribution lies precisely in identifying these different coordination alternatives for actors, that can be all present in a network to improve its functioning.

The coordination alternatives for actors imply the existence of actor coordination classes, that can be assigned various semantics. Assume that a direct actor-actor coordination link is established among two non-failed actors via an intermediary, actor failed. This can happen only if the two are in range which, given their previous indirect communication via the intermediary, is not always the case. However, when this happens, it means that the two actors are rather powerful devices with large enough ranges. Semantically, we can define a subset of the actor set in a WSAN which are strategic to the network and are so powerful that can communicate directly with each other even if intermediary actors fail. The most security-sensitive information and the real-time constrained information can thus be transmitted via links among such strategic actors. Indirect links between actors can be used to transmit non-sensitive information. The fact that indirect coordination among any two actors can be established through sensors provides a fault tolerance support for actor coordination. If the direct actor links are enabled then the communication takes place via them, but if not then a `rescue' route can be established via the sensor infrastructure. Thus, the sensor nodes provide the backup infrastructure on which the actor coordination can rely.

In order to prove the local path existence property, we employ the Event-B formal method for constructing a new actor path. Event-B~\cite{6,7} is an extension of the B formalism~\cite{bbook} for specifying distributed and reactive systems. A system model is gradually specified on several levels of abstraction, always ensuring that a more concrete model is a \emph{correct implementation} of an abstract model. The language and proof theory of Event-B are based on logic and set theory. The correctness of the stepwise construction of formal models is ensured  by discharging a set of proof obligations: if these obligations hold, then the development is mathematically shown to be correct. Event-B comes with the associated tool RODIN~\cite{7,8}, which automatically discharges part of the proof obligations and also provides the means for the user to discharge interactively the remaining proofs.

The paper is organized as follows. In Section 2 we briefly overview the Event-B formalism and present the recovery algorithm. In Section 3 we model the direct actor links recovery mechanism and in Section 4 the indirect actor links recovery. In Section 5 we model the sensor infrastructure and the indirect actor links recovery via the sensors. In Section 6 we present the proof statistics of our model and in Section 7 we further discuss about the contribution of our paper. In Section 8 we briefly conclude the paper.

\section{Preliminaries}
This section briefly overviews our modeling formalism Event-B and also describes the recovery algorithm to be modeled in this paper.

\paragraph{\bf{Event-B}}
Each Event-B model consists of two components called \emph{context} and \emph{machine}. A context describes the static part of the model, i.e., it introduces new types and constants. The properties of these types and constants are gathered as a list of axioms. A machine represents the dynamic part of the model, consisting of variables that define the \emph{state} of the model and operations called \emph{events}. The structure of an Event-B machine is given in Fig.~\ref{Figure2}. The system properties that should be preserved during the execution are formulated as a list of \emph{invariant} predicates over the \emph{state} of the model.

An event, modeling state changes, is composed of a \emph{guard} and an \emph{action}. The guard is the necessary condition under which an event might occur; if the guard holds, we call the event \emph{enabled}. The action determines the way in which the state variables change when the event occurs. For initializing the system, a sequence of actions is defined. When the guards of several events hold at the same time, then only one event is non-deterministically chosen for execution. If some events have no variables in common and are enabled at the same time, then they can be considered to be executed in parallel since their sequential execution in any order gives the same result.

\begin{figure}
 \label{Figure2}
 \centering
  \begin{small}
  \begin{center}
   \frame{\parbox{300pt}{\smallskip\hspace{5pt}\textbf{MACHINE} machine-name

	\hspace{20pt}\textbf{VARIABLES} list of variables

	\hspace{20pt}\textbf{INVARIANTS} list of invariants/predicates

	\hspace{20pt}\textbf{EVENTS}

	\hspace{30pt}\textbf{INITIALISATION}

	\hspace{30pt}BEGIN

	\hspace{40pt} list of actions

	\hspace{30pt}END

	\hspace{30pt}\textbf{event-name}

	\hspace{30pt}WHEN

	\hspace{40pt} list of guards

	\hspace{30pt}THEN

	\hspace{40pt} list of actions

	\hspace{30pt}END	

	\hspace{5pt}\textbf{END}\smallskip
}}\end{center}\end{small}
\caption{MACHINE definition in Event-B}
 \end{figure}

A model is developed by a number of correctness preserving steps called \emph{refinements}. One  form of model refinement can add new data and new behavior events on top of the already existing data and behavior but in such a way that the introduced behavior does not contradict or take over the abstract machine behavior. In addition to this \emph{superposition} refinement~\cite{katzSuperimp} we may also use other refinement forms, such as \emph{algorithmic} refinement~\cite{mpc89}. In this case, an event of an abstract machine can be refined by several corresponding events in a refined machine. This will model different branches of execution, that can for instance take place in parallel and thus can improve the algorithmic efficiency.

\paragraph{\bf{The recovery algorithm}}
In this algorithm, the detection of a failed node leads to the communication links among non-failed actor nodes to be reconstructed via sensor nodes. The mechanism has three parts: detecting a failed actor, selecting the shortest path, and establishing the selected path through sensor nodes. When actor neighbors of an actor node do not receive any acknowledgment from that actor node, they detect it as failed. At this time, the neighbors of the failed node have to investigate whether this failure has produced separated partitions. If there is no partitioning, then nothing is done except updating the neighbor lists in nodes. However, if there are some separated partitions, a new path should be selected and established.

We assume that each actor node has information about its immediate neighbors (1-hop neighbors) and 2-hop neighbors (the neighbors of the neighbors). Based on this information, the non-failed actors can recover their communication links upon detecting a failed (intermediary) actor. These links are formed based on the node \emph{degree} information (the number of immediate neighbors) and on the relative distance between actor nodes. In this paper we focus only on the self-recovery mechanism via sensors, due to lack of space. The complete model of the algorithm can be found in~\cite{TR}.

Our formalization shows that, given networks of sensors and actors, the actors can \emph{always} reconstruct coordination links between themselves, by using local information and sensors as intermediate nodes. In order to prove this property, we model the network at three increasing levels of detail so that each model is a refinement of the previous one. In the initial model, we specify the actor network and the recovery mechanism of the direct actor links. In the second model, we add new data and events to model the list of 1-hop and 2-hop neighbors for every node and model the recovery via indirect actor links. In the third model, we distinguish among sensor and actor nodes and their corresponding networks and model the recovery via the sensor infrastructure. In the following, we describe these three recovery models.


\section{Recovery via Direct Actor Links}



The context of our initial model contains the definition of constants and sets as well as our model assumptions as axioms. A finite (axiom 6) and non-empty (axiom 7), generic set $NODE$ describes all the network nodes. 
These nodes can be either sensor nodes or actor nodes, hence the set $NODE$ is partitioned into the \emph{sensors} and \emph{actors} sets, where \emph{sensors} and \emph{actors} are predefined sets of sensors and actors respectively (axiom 8).
 $STATUS$ denotes the set $\{ok, fail\}$, where the constant $fail$ stands for a failed node and the constant $ok$ for a non-failed node (axiom 1). We also define the $closure$ constant that models the transitive closure of a binary relation on the set $NODE$ (axioms 2-5). We use this constant in order to dynamically construct all the possible paths for the current network.

\begin{small}

\begin{center}
\frame{\parbox{300pt}{\smallskip

\hspace{5pt}\textbf{constants} $  closure  \ ok \ fail$

\hspace{5pt}\textbf{sets} $NODE \ STATUS \ sensors \ actors$

\hspace{5pt}\textbf{axioms}
	
\hspace{20pt}\textbf{@axm1} $partition(STATUS, \{ok\}, \{fail\})$

\hspace{20pt}\textbf{@axm2} $closure \in(NODE\leftrightarrow NODE)\rightarrow (NODE\leftrightarrow NODE)$

\hspace{20pt}\textbf{@axm3} $\forall r \cdot r \subseteq closure(r)$

\hspace{20pt}\textbf{@axm4} $\forall r \cdot closure(r);r\subseteq closure(r)$

\hspace{20pt}\textbf{@axm5} $\forall r,s \cdot r \subseteq s \wedge s;r \subseteq s  \Rightarrow closure(r)\subseteq s$

\hspace{20pt}\textbf{@axm6} $finite(NODE)$

\hspace{20pt}\textbf{@axm7} $NODE\neq \emptyset $

\hspace{20pt}\textbf{@axm8} $partition(NODE, sensors, actors)$\smallskip}}\end{center}\end{small}

\noindent In the machine part of our initial model we have six events and six invariants as shown below. The status of each node (non-failed or failed) is modeled with the function $Status$ mapping each node in $NODE$ to $ok$ or $fail$ (invariant 1). The relation $ANET$ denotes the bidirectional, non-failed actor links (invariant 2 and 6). This relation is non-reflexive (invariant 4), expressed with the domain restriction operator $\triangleleft$ and symmetric (invariant 5), expressed with the relation inverse operator $\sim$. This means that an $ANET$ link from a node to itself is prohibited and if node $a$ has a link with node $b$, then node $b$ also has a link with the node $a$. The set $FailedNodeNeigh$ denotes non-failed actors (invariant 7). This set is updated when a node is detected as failed. We also model that the network is active continuously with theorem {\bf THM1} that ensures that, always, at least one event is enabled (i.e., the disjunction of all the events guards is true). We express this constraint with a theorem instead of an invariant for technical reasons detailed in~\cite{TR}.


\begin{small}
\begin{center}
\frame{\parbox{350pt}{\smallskip\hspace{5pt}\textbf{INVARIANTS}

\hspace{20pt}\textbf{@inv1}  $Status\in NODE\rightarrow STATUS$

\hspace{20pt}\textbf{@inv2}  $ANET\in actors \leftrightarrow actors$

\hspace{20pt}\textbf{@inv3}  $FailedNodeNeigh\subseteq actors$

\hspace{20pt}\textbf{@inv4}  $actors\triangleleft id \cap ANET=\emptyset$

\hspace{20pt}\textbf{@inv5}  $ANET = ANET\sim$

\hspace{20pt}\textbf{@inv6} $\forall n,m \cdot  n\mapsto m\in ANET \Rightarrow Status(n)= ok \wedge Status(m)= ok$

\hspace{20pt}\textbf{@inv7} $FailedNodeNeigh\subseteq (Status\sim[\{ok\}])$

		

%

\hspace{20pt}\textbf{theorem @THM1}$(\exists n \cdot Status(n)= fail \ \wedge \ n \in actors \ \wedge \ FailedNodeNeigh= \emptyset)$

\hspace{50pt}$\vee \ (\exists \ n,m \cdot Status(n)= ok \ \wedge \ Status(m)= ok \ \wedge \ n\mapsto m\notin ANET$

\hspace{55pt}$\ \ \ \quad \quad \quad \wedge \ n \in actor \ \wedge \ m \in actor \ \wedge \ n\neq m \ \wedge \ FailedNodeNeigh= \emptyset)$

\hspace{50pt}$\vee \ (\exists \ n \cdot Status(n)= ok\ \wedge \ n \in Actors \ \wedge \ FailedNodeNeigh= \emptyset)$

\hspace{50pt}$\vee \ (\exists \ n,k \cdot n \in FailedNodeNeigh \ \wedge \ k \in FailedNodeNeigh$

\hspace{55pt}$\quad \quad \quad \ \ \wedge \ n\mapsto k\notin closure(ANET))$

\hspace{50pt}$\vee  (\exists \ n,k \cdot n \in FailedNodeNeigh \ \wedge \ k \in FailedNodeNeigh$

\hspace{55pt}$\quad \quad \quad \ \wedge \ n\mapsto k\in closure(ANET))$\smallskip


}}\end{center}\end{small}

\noindent The initialisation event sets the status of all the nodes to $fail$; therefore, the $ANET$ relation should be empty based on invariant 6. The set $FailedNodeNeigh$ is set to $\emptyset$ because it is a sub-set of non-failed nodes (invariant 7)  and there are no non-failed nodes in the network at initialization.

Except initialisation, the events in the initial model add actor nodes ({\bf AddNode}) and actor links ({\bf AddLink}), remove actor nodes and their corresponding links ({\bf RemoveNode}) and also abstractly recover connections when an actor fails ({\bf FaultDetRec} and {\bf FaultDetRec2}).

\begin{small}
\begin{center}
\frame{\parbox{200pt}{\smallskip\hspace{5pt}\textbf{INITIALISATION}

\hspace{20pt}\textbf{then}

\hspace{30pt}\textbf{@act1} $Status:= NODE\times \{fail\}$

\hspace{30pt}\textbf{@act2} $ANET:= \emptyset$

\hspace{30pt}\textbf{@act3} $FailedNodeNeigh := \emptyset$
\smallskip
}}\end{center}\end{small}

\noindent In the {\bf AddNode} event, every actor that is added overwrites the function $Status$.

\begin{small}
\begin{center}
\frame{\parbox{200pt}{\smallskip\hspace{5pt}\textbf{AddNode}

\hspace{20pt}\textbf{any} n \textbf{where}

\hspace{30pt}\textbf{@grd1} $Status(n)=fail$

\hspace{30pt}\textbf{@grd2} $FailedNodeNeigh = \emptyset$

\hspace{30pt}\textbf{@grd3} $n \in actors$

\hspace{20pt}\textbf{then}

\hspace{30pt}\textbf{@act1} $Status(n):= ok$

\hspace{20pt}\textbf{end}\smallskip
}}\end{center}\end{small}


\noindent In the {\bf AddLink} event we add a link in both directions, to meet invariant~4.

\begin{small}
\begin{center}
\frame{\parbox{270pt}{\smallskip\hspace{5pt}\textbf{AddLink}

\hspace{20pt}\textbf{any} n m \textbf{where}

\hspace{30pt}\textbf{@grd1} $Status(n)= ok \wedge Status(m)=ok$

\hspace{30pt}\textbf{@grd2} $n\mapsto m \notin ANET$

\hspace{30pt}\textbf{@grd3} $n\neq m$

\hspace{30pt}\textbf{@grd4} $n \in actors \wedge m \in actors$

\hspace{30pt}\textbf{@grd5} $FailedNodeNeigh = \emptyset$

\hspace{20pt}\textbf{then}

\hspace{30pt}\textbf{@act1} $ANET:=ANET\cup \{n\mapsto m, m\mapsto n\}$

\hspace{20pt}\textbf{end}\smallskip
}}\end{center}\end{small}

\noindent The {\bf RemoveNode} event changes the status of an actor from $ok$ to $fail$; also, all the links of that actor are removed from $ANET$, expressed with the domain substraction operator $\domsub$ and the range substraction operator $\ransub$. In addition, neighbors of that actor become members of the $FailedNodeNeigh$ set. This means that we model the situation where one actor fails at a time, because {\bf RemoveNode} is not enabled again until $FailedNodeNeigh$ becomes empty again (guard 3). Although restrictive, we choose this approach in this paper for simplicity. We observe that even if the network is left with one or zero non-failed actors, our algorithm does not deadlock because the {\bf AddNode} event is still enabled.

\begin{small}
\begin{center}
\frame{\parbox{220pt}{\smallskip\hspace{5pt}\textbf{RemoveNode}

\hspace{20pt}\textbf{any} $n$ \textbf{where}

\hspace{30pt}\textbf{@grd1} $Status(n)= ok$

\hspace{30pt}\textbf{@grd2} $n \in actors$

\hspace{30pt}\textbf{@grd3} $FailedNodeNeigh = \emptyset$

\hspace{20pt}\textbf{then}

\hspace{30pt}\textbf{@act1} $Status(n):= fail$

\hspace{30pt}\textbf{@act2} $ANET:= \{n\}\domsub ANET\ransub \{n\}$

\hspace{30pt}\textbf{@act3} $FailedNodeNeigh := ANET[\{n\}]$

\hspace{20pt}\textbf{end}\smallskip
}}\end{center}\end{small}


\noindent Removing an actor from the network can lead to separated network partitions. The event {\bf FaultDetRec} detects whether removing an actor has created separated partitions or not. If two neighbors of a failed actor had no connection through other actors (i.e., there was no path from one node to the other, expressed by guard 2 of {\bf FaultDetRec}), then a partition is formed. To recover from this partitioning of communication, a direct actor-actor link is established (act 1). As $FailedNodeNeigh$ is a subset of a finite set (invariant 6, invariant 1, axiom 6), we observe that the {\bf FaultDetRec} event can be enabled only a finite number of times, hence the recovery operation terminates. Technically, this is true because $card(FailedNodeNeigh)$ decreases at each execution of {\bf FaultDetRec} and eventually the guard of {\bf FailedDetRec} will hold no longer. We observe that $n \neq k$ due to guard 2 (the $closure$ construct is reflexive).

\begin{small}
\begin{center}
\frame{\parbox{270pt}{\smallskip\hspace{5pt}\textbf{FaultDetRec}

\hspace{20pt}\textbf{any} n k \textbf{where}

\hspace{30pt}\textbf{@grd1} $n \in FailedNodeNeigh \wedge k \in FailedNodeNeigh$



\hspace{30pt}\textbf{@grd2} $n\mapsto k \notin closure(ANET)$

\hspace{20pt}\textbf{then}

\hspace{30pt}\textbf{@act1} $ANET:=ANET\cup\{n\mapsto k,k\mapsto n\} $

\hspace{30pt}\textbf{@act2} $FailedNodeNeigh := FailedNodeNeigh \setminus \{n\} $

\hspace{20pt}\textbf{end}\smallskip
}}\end{center}\end{small}


\noindent The {\bf FaultDetRec2} event treats the situation when a failure is detected but an alternative path already exists between the neighbors of the failed actor ($n\mapsto k\in closure(ANET)$). In this case, $FailedNodeNeigh$ is simply updated. We observe that in the case $n= k$, the last element of $FailedNodeNeigh$ is removed by this event.

\begin{small}
\begin{center}
\frame{\parbox{270pt}{\bigskip\hspace{5pt}\textbf{FaultDetRec2}

\smallskip\hspace{20pt}\textbf{any} n k \textbf{where}

\hspace{30pt}\textbf{@grd1} $n \in FailedNodeNeigh \wedge k \in FailedNodeNeigh$

\hspace{30pt}\textbf{@grd2} $n\mapsto k \in closure(ANET)$

\hspace{20pt}\textbf{then}

\hspace{30pt}\textbf{@act2} $FailedNodeNeigh := FailedNodeNeigh \setminus \{n\}$

\hspace{20pt}\textbf{end}\smallskip
}}\end{center}\end{small}

\noindent Overall, the initial model presented in this section describes the non-deterministic addition and removal of actor nodes and actor links in a dynamic (wireless sensor-actor) network for whom the network partitioning problem can be detected and recovered from via direct actor links. The recovery assumes some global network knowledge for the recovery, expressed by $closure(ANET)$. Also, the recovery mechanism establishes direct links among the non-failed actor neighbors of the failed actor. Both the recovery assumption and the recovery mechanism can be used in practice only for \emph{strategic actors}, i.e., actors whom range is sufficiently large to check $closure(ANET)$ and establish direct actor links. The following model considers more localized assumptions as well as indirect recovery paths.

\section{Recovery via Indirect Actor Links}

\noindent In the previous model we have considered the actor network able to access knowledge about itself while in the model in this section we assume that each actor has access only to information of its 1-hop neighbors and 2-hop neighbors. We now refine the initial model and define a new relation $l\_net$ that, for each actor, keeps track of the 1-hop and 2-hop neighbors. The relation $l\_net$ relates three nodes as defined by invariant 1 below and is not reflexive, modeled by invariant 2 below. The meaning of this relation is that a 1-hop neighbor $m$ of a node $n$ is denoted by $n\mapsto m\mapsto m \in l\_net$ and a 2-hop neighbor $m$ of a node $n$ is denoted by $n\mapsto m\mapsto k\in l\_net$. In the first example, $m$ is locally related to $n$ via $m$ (itself, i.e., via a direct link) and in the second example $m$ is locally related to $n$ via $k$ (i.e., $m$ is a 2-hop neighbor of $n$, while $k$ is a 1-hop neighbor of $n$). The relation $l\_net$ describes all these \emph{localized} links between nodes. The goal of this refinement step is to supplement the global knowledge of the network in the initial model (via $closure(ANET)$) with a localized knowledge formalized with the relation $l\_net$.

\begin{small}
\begin{center}
\frame{\parbox{330pt}{

\smallskip\hspace{5pt}\textbf{@inv1} $l\_net\in actors \times actors \leftrightarrow NODE $

\hspace{5pt}\textbf{@inv2} $actors \triangleleft id \cap dom(l\_net)=\emptyset$ \smallskip
}}\end{center}\end{small}

\noindent When a new link is added between two actors the $l\_net$ relation also needs to be updated. Therefore, the {\bf AddlLink} event is extended to also add links to $l\_net$. For every two nodes $n$ and $m$ which have a direct link, $n \mapsto m\mapsto m$ and $m \mapsto n \mapsto n$ are added, meaning that $n$ has a link with $m$ through $m$ ($m$ is a 1-hop neighbor $n$) and $m$ has a link with $n$ through $n$ ($n$ is a 1-hop neighbor of $m$).

\begin{small}
\begin{center}
\frame{\parbox{280pt}{\smallskip\hspace{5pt}\textbf{AddLink}

\hspace{20pt}\textbf{extends} AddLink

\hspace{20pt}\textbf{then}

\hspace{30pt}\textbf{@act2} $l\_net:=l\_net\cup\{n\mapsto m\mapsto m,m\mapsto n\mapsto n\}$

\hspace{20pt}\textbf{end}\smallskip
}}\end{center}\end{small}

\noindent The {\bf Addl\_net2hopLink} event is a newly introduced event that handles the addition of 2-hop neighbor links for actors. If an actor has a direct link with two other actors, then these actors will be 2-hop neighbors of each other:

\begin{small}
\begin{center}
\frame{\parbox{300pt}{\smallskip\hspace{5pt}\textbf{Addl\_net2hopLink}

\hspace{20pt}\textbf{any} n m k \textbf{where}

\hspace{30pt}\textbf{@grd1} $Status(n)= ok \wedge Status(m)= ok \wedge Status(k)= ok$

\hspace{30pt}\textbf{@grd2} $m\mapsto k\mapsto k\in l\_net \wedge n\mapsto m\mapsto m\in l\_net \wedge $

\hspace{65pt} $n\mapsto k\mapsto m\notin l\_net \wedge k\mapsto n\mapsto m\notin l\_net$

\hspace{30pt}\textbf{@grd3} $m\neq n \wedge n\neq k \wedge m\neq k$

\hspace{30pt}\textbf{@grd4} $FailedNodeNeigh = \emptyset$

\hspace{20pt}\textbf{then}

\hspace{30pt}\textbf{@act1} $l\_net:=l\_net\cup\{n\mapsto k\mapsto m,k\mapsto n\mapsto m\}$

\hspace{20pt}\textbf{end}\smallskip
}}\end{center}\end{small}

\noindent When removing an actor node, all its connections should be removed. Thus, in the {\bf RemoveNode} event a new action is added which removes all the immediate links with the failed actor in the $l\_net$ relation. 
The expression $\{n\}\times dom(ANET)\times dom(ANET)$ describes all the links of $n$, either direct connections (1-hop neighbors) or indirect connections (2-hop neighbors) and the expression $dom(ANET)\times \{n\}\times\{n\}$ describes all the links between immediate neighbors of $n$ and $n$.

\begin{small}
\begin{center}
\frame{\parbox{300pt}{\smallskip\hspace{5pt}\textbf{RemoveNode}

\hspace{20pt}\textbf{extends} RemoveNode

\hspace{20pt}\textbf{then}

\hspace{30pt}\textbf{@act4} $l\_net:= l\_net\setminus ((\{n\}\times dom(ANET)\times dom(ANET)) \cup $

\hspace{65pt}$(dom(ANET)\times \{n\}\times\{n\}))$

\hspace{20pt}\textbf{end}\smallskip
}}
\end{center}
\end{small}

\noindent We now need to model the detection of failed nodes and the recovery of links based on local information instead of being based on all the network topology as described by $closure(ANET)$. We now use $l\_net$ information in addition to $ANET$ for detecting an actor failure (guard 3) and recovering links in the {\bf FaultDetRec} event.

\begin{small}
\begin{center}
\frame{\parbox{350pt}{\smallskip\hspace{5pt}\textbf{FaultDetRec}

\hspace{20pt}\textbf{extends} FaultDetRec

\hspace{20pt}\textbf{any} n m k  \textbf{where}

%
%
\hspace{30pt}\textbf{@grd3} $n\mapsto k\mapsto m\in l\_net \wedge n\mapsto m\mapsto m\notin l\_net \wedge$

\hspace{65pt} $k\mapsto n\mapsto m\in l\_net \wedge k\mapsto m\mapsto m\notin l\_net$
%


\hspace{20pt}\textbf{then}

%
\hspace{30pt}\textbf{@act3} $l\_net:\mid l\_net' \subseteq (l\_net\setminus (\{n\mapsto k\mapsto m, k\mapsto n\mapsto m\}\cup$

\hspace{65pt}$(ANET[\{n\}]\times\{m\}\times\{n\})\cup(ANET[\{k\}]\times\{m\}\times\{k\})))$

\hspace{65pt}$\cup(ANET[\{k\}]\times\{n\}\times\{k\})\cup (\{n\}\times ANET[\{k\}]\times\{k\})$

\hspace{65pt}$\cup(ANET[\{n\}]\times\{k\}\times\{n\})\cup(\{k\}\times ANET[\{n\}]\times\{n\})\cup$

\hspace{65pt}$(\{k\}\times\{n\}\times(NODE\setminus\{m\}))\cup(\{n\}\times\{k\}\times(NODE\setminus\{m\}))$\smallskip
}}\end{center}\end{small}

\noindent When actor $m$ is detected as failed, neighbors of $m$ ($n$ and $k$) that have a connection with each other through $m$ ($n\mapsto k\mapsto m$ and $k\mapsto n\mapsto m$) need to find an alternative path toward each other. If there is no other route in $ANET$ ($n\mapsto k\notin closure(ANET)$), then $l\_net$ should be updated in two phases, by removing expired links and adding new routes. Since $m$ is failed, links between $n$ and $k$ through $m$ are not anymore valid, so $n\mapsto k\mapsto m$ and $k\mapsto n\mapsto m$ are removed from $l\_net$. In addition, links describing the immediate neighbors of $n$ ($ANET[\{n\}]$) and of $k$ ($ANET[\{k\}]$) to $m$ via $n$ and $k$, respectively, are removed from $l\_net$. The second phase of the updating process is adding new links to connect $n$ and $k$. In this refinement, since we still have no information about sensors, we define that actor $n$ can establish a link with actor $k$ through any node except $m$ which is failed: $\{n\}\times\{k\}\times(NODE\setminus\{m\})$ and similarly for actor $k$ to establish a new link with actor $n$: $\{k\}\times\{n\}\times(NODE\setminus\{m\})$. When node $n$ establishes a link with $k$, neighbors of $n$ also need to add node $k$ to their 2-hop neighbors list ($ANET[\{n\}]\times\{k\}\times\{n\}$). Moreover, neighbors of $k$ need to add $n$ to their 2-hop neighbors list ($ANET[\{k\}]\times\{n\}\times\{k\}$). The updating process of $l\_net$ is described by action 2 in the {\bf FaultDetRec} event.

\noindent We add a new action to event {\bf FaultDetRec2} that updates $l\_net$ by removing all the links with the failed actor or through it.



\begin{small}
\begin{center}
\frame{\parbox{320pt}{\smallskip\hspace{5pt}\textbf{FaultDetRec2}

\hspace{20pt}\textbf{extends} FaultDetRec2

\hspace{20pt}\textbf{any} n m k \textbf{where}



\hspace{30pt}\textbf{@grd3} $n\mapsto k\mapsto m\in l\_net \wedge n\mapsto m\mapsto m\notin l\_net \wedge$

\hspace{65pt} $k\mapsto n\mapsto m\in l\_net \wedge k\mapsto m\mapsto m\notin l\_net$


\hspace{20pt}\textbf{then}

\hspace{30pt}\textbf{@act2} $l\_net:=l\_net\setminus(\{n\mapsto k\mapsto m,k\mapsto n\mapsto m\}\cup$

\hspace{65pt}$(ANET[\{n\}]\times\{m\}\times\{n\})\cup(ANET[\{k\}]\times\{m\}\times\{k\}))$\smallskip

}}\end{center}\end{small}

We observe that $l\_net$ is an elegant data structure relating two actor nodes in its domain via a third node in its range. The model described in this section is a formal refinement of the one in the previous section. This means that the old invariants still hold for the extended model, in addition to the two new ones. Moreover, the link recovery terminates in a finite number of steps. Indirect links between actors are now established non-deterministically based on localized information. These types of links can be further refined to a more deterministic form, as we show in the following section.

\section{Sensor-Based Recovery}

In the previous model, we have defined $l\_net$ as a subset of actor relations which after a failure detection non-deterministically is upadeted due to lack of knowledge about sensor nodes. In this model, we add sensor nodes and specify more concretely how replacement links through sensors are added after detecting an actor failure. We introduce two new relations on $NODE$ (invariant 1 and invariant 2), $SNET$ and $SANET$, the former representing links among sensor nodes and the latter depicting links between sensor and actor nodes.

\begin{small}
\begin{center}
\frame{\parbox{320pt}{\smallskip\hspace{5pt}

\hspace{5pt}\textbf{@inv1} $SNET\in sensors \leftrightarrow sensors$

\hspace{5pt}\textbf{@inv2} $SANET\in NODE\leftrightarrow NODE$

\hspace{5pt}\textbf{@inv3} $SNET\cap ANET=\emptyset$

\hspace{5pt}\textbf{@inv4} $ANET\cap SANET=\emptyset$

\hspace{5pt}\textbf{@inv5} $SNET\cap SANET=\emptyset$

\hspace{5pt}\textbf{@inv6} $SNET=SNET\sim$

\hspace{5pt}\textbf{@inv7} $SANET=SANET\sim$

\hspace{5pt}\textbf{@inv8} $sensors\triangleleft id \cap SNET=\emptyset$

\hspace{5pt}\textbf{@inv9} $NODE\triangleleft id \cap SANET=\emptyset$

\hspace{5pt}\textbf{@inv10} $\forall n,m \cdot n\mapsto m\in SANET \Rightarrow$

\hspace{45pt} $(n \in actors \wedge m \in sensors)$

\hspace{45pt} $\vee (m \in actors \wedge n \in sensors)$

\hspace{5pt}\textbf{@inv11} $\forall n,m\cdot n\mapsto m\in SNET\Rightarrow Status(n)= ok \wedge Status(m)= ok$

\hspace{5pt}\textbf{@inv12} $\forall n,m\cdot n\mapsto m\in SANET\Rightarrow Status(n)=ok \wedge Status(m)= ok$




\hspace{5pt}\textbf{@inv13} $\forall n,k,x,y\cdot n\mapsto k\mapsto x\in l\_net \wedge k\mapsto n\mapsto y\in l\_net \wedge$

\hspace{45pt} $x \in sensors \wedge y \in sensor \Rightarrow$

\hspace{45pt} $ x\in SANET[\{n\}] \wedge\ y\in SANET[\{k\}] \wedge x\mapsto y\in closure(SNET)$ \smallskip

}}\end{center}\end{small}

\noindent These relations describe links between nodes at a different level, hence they are disjoint from the actor links modeled by $ANET$ (invariant 3 and invariant 4). $SNET$ and $SANET$ are also disjoint sets (invariant 5). Moreover, they are symmetric and non-reflexive sets as shown by invariants 6-9.  We also formalize that for each link $n\mapsto m$ in $SANET$ one of these nodes should be a sensor node and the other one should be an actor node (invariant 10). The next two invariants (invariant 11 and 12) model that every node of a link in either $SNET$ or $SANET$ should be non-failed. 
Invariant 13 models that if there is a link between two actor nodes via sensor nodes in $l\_net$, the involved sensor nodes are within the range of $l\_net$, the respective actor-sensor links belong to $SANET$ and the sensors themselves have at least one path toward each other within $closure(SNET)$.


In the previous model, removing an actor and all its connections was modeled by the {\bf RemoveNode} event. In this model we refine {\bf RemoveNode} by adding a new action for updating $SANET$ after removing an actor node (action 5). Also, all connections through sensor nodes towards a failed node should be removed from $l\_net$ (action 4). 

\begin{small}
\begin{center}
\frame{\parbox{360pt}{\smallskip\hspace{5pt}\textbf{RemoveNode}

\hspace{20pt}\textbf{refines} RemoveNode

%

\hspace{30pt}\textbf{then}

%
%
\hspace{30pt}\textbf{@act4} $l\_net:=l\_net\setminus((\{n\}\times dom(ANET)\times dom(ANET) )\ \cup $

\hspace{60pt}$(dom(ANET)\times\{n\}\times\{n\})\cup (dom(ANET)\times\{n\}\times dom(SNET)))$

\hspace{30pt}\textbf{@act5} $SANET:=\{n\}\domsub SANET\ransub \{n\}$

\hspace{20pt}\textbf{end}\smallskip

}}\end{center}\end{small}


%
%
%
%
%
%

%
%
%
%
%

\noindent In this model we have two new events for adding links between sensor nodes in $SNET$ and links between sensor and actor nodes in $SANET$: {\bf AddSLink} and {\bf AddSALink}. \\

\begin{small}
\begin{center}
\frame{\parbox{300pt}{\smallskip\hspace{5pt}\textbf{AddSLink}

\hspace{20pt}\textbf{any} n m \textbf{where}

\hspace{30pt}\textbf{@grd1} $Status(n)= ok \wedge Status(m)= ok$


\hspace{30pt}\textbf{@grd2} $n\mapsto m\notin SNET$

\hspace{30pt}\textbf{@grd3} $n\neq m$

\hspace{30pt}\textbf{@grd4} $n \in sensors \wedge m \in sensors$

\hspace{20pt}\textbf{then}

\hspace{30pt}\textbf{@act1} $SNET:=SNET\cup\{n\mapsto m,m\mapsto n\}$

\hspace{20pt}\textbf{end}\smallskip
}}\end{center}\end{small}

\begin{small}
\begin{center}
\frame{\parbox{330pt}{\smallskip\hspace{5pt}\textbf{AddSALink}

\hspace{20pt}\textbf{any} n m \textbf{where}

\hspace{30pt}\textbf{@grd1} $Status(n)= ok \wedge Status(m)= ok$

\hspace{30pt}\textbf{@grd2} $(n \in actors \wedge m \in sensors$

\hspace{45pt} $\vee (n \in sensors \wedge m \in actors)$

\hspace{30pt}\textbf{@grd3} $n\mapsto m\notin SANET$

\hspace{30pt}\textbf{@grd4} $n\neq m$

\hspace{20pt}\textbf{then}

\hspace{30pt}\textbf{@act1} $SANET:=SANET\cup\{n\mapsto m,m\mapsto n\}$

\hspace{20pt}\textbf{end}\smallskip
}}\end{center}\end{small}

\noindent The {\bf AddSLink} event is similar to {\bf AddLink} with a different guard that models that, for every map $n\mapsto m$ added in {\bf AddSLink}, $n$ and $m$ should be sensor nodes. The {\bf AddSALink} event is for adding links between sensor and actors.

\noindent The event {\bf FaultDetRec} which models the recovery mechanism after an actor failure is refined using information of $SNET$ and $SANET$. Compared to the previous version of the event, there are two additional parameters $x$, $y$ as sensor nodes which have connections with actor nodes $n$ and $k$, respectively. Also, $x$ and $y$ have either a direct link or an indirect one towards each other, via $closure(SNET)$. Moreover, the actors $n$ and $k$ have no connection with each other (guard 7).

\begin{small}
\begin{center}
\frame{\parbox{350pt}{\smallskip\hspace{5pt}\textbf{FaultDetRec}

\hspace{20pt}\textbf{refines} FaultDetRec

\hspace{20pt}\textbf{any} n m k x y \textbf{where}

%
%
%
%
%

\hspace{30pt}\textbf{@grd4} $x\in SANET[\{n\}] \wedge y\in SANET[\{k\}]$

\hspace{30pt}\textbf{@grd5} $x\mapsto y\in closure(SNET)$

\hspace{30pt}\textbf{@grd6} $m \in actors$

\hspace{30pt}\textbf{@grd7} $n\mapsto k\notin dom(l\_net\setminus\{n\mapsto k\mapsto m\})$

\hspace{20pt}\textbf{then}


\hspace{30pt}\textbf{@act3} $l\_net:=(l\_net\setminus(\{n\mapsto k\mapsto m, k\mapsto n\mapsto m\}$

\hspace{70pt}$\cup(ANET[\{n\}]\times\{m\}\times\{n\})\cup(ANET[\{k\}]\times\{m\}\times\{k\})))$

\hspace{70pt}$\cup(ANET[\{k\}]\times\{n\}\times\{k\})\cup (\{n\}\times ANET[\{k\}]\times\{k\})$

\hspace{70pt}$\cup(ANET[\{n\}]\times\{k\}\times\{n\})\cup(\{k\}\times ANET[\{n\}]\times\{n\})$

\hspace{70pt}$\cup \{n\mapsto k\mapsto x, k\mapsto n\mapsto y\}$

\hspace{20pt}\textbf{end}\smallskip

}}\end{center}\end{small}

\noindent The action 3 in {\bf FaultDetRec} was non-deterministic in the previous model. We now refine this assignment to a deterministic one. We replace $\{k\}\times \{n\}\times NODE\setminus\{m\}$ with $k\mapsto n\mapsto y$ and similarly $\{n\}\times \{k\}\times NODE\setminus\{m\}$ is replaced with  $n\mapsto k\mapsto x$.\\

\noindent The action in the {\bf FaultDetRec2} event is unchanged. However, we strengthen the guard of the event by adding guard 6 that guarantees the existence of a link between two direct neighbors of a failed node via other nodes than the failed one.\\

\begin{small}
\begin{center}
\frame{\parbox{250pt}{\smallskip\hspace{5pt}\textbf{FaultDetRec2}

\hspace{20pt}\textbf{refines} FaultDetRec2

\hspace{20pt}\textbf{where}

\hspace{30pt}\textbf{@grd6} $n\mapsto k\in dom(l\_net\setminus\{n\mapsto k\mapsto m\})$

\hspace{20pt}\textbf{end}\smallskip

}}\end{center}\end{small}

In this third model we uncover the sensor infrastructure and employ it for the actor recovery. This model is a refinement of the previous models, respectiing all the introduced invariants, old and new. The recovery is terminating in a finite number of steps as for the previous two models. The third model illustrates the usage of sensors as a fault tolerance mechanism for the actor coordination.

\paragraph{{\bf An Additional Model}}

In the model presented in this section, we re-establish connections (through sensor nodes) between pairs of actors which were direct neighbors of a failed actor node. However, this is not an optimal mechanism since actor nodes can be far from each other. In this case their reestablished connections would involve numerous sensor nodes, while there might be a shorter path for this. To determine the shortest path between these actor nodes we need to introduce information about the physical location of the nodes. We describe this refinement step in~\cite{TR} in all the details but skip it here due to lack of space.

\section{Proof Statistics}

\noindent The proof statistics of our development is shown in Table~\ref{Proof Statistics}. These figures express the number of proof obligations generated by the Rodin Platform as well as the number of  obligations automatically discharged by the platform and those interactively proved. There are significantly more proof obligations in the second refinement due to introducing the details of sensor networks and refining the recovery algorithm to use the sensor nodes. In order to guarantee the correctness of the recovery algorithm, new invariants had to be added and proven. Due to the lack of adequate automatic support in the Rodin platform for reasoning about set comprehension and unions, we faced with a high number of interactive proofs. In addition, the interactive proving often involved manually suggesting values to discharging various properties containing logical disjunctions or existential quantifiers. Another proving difficulty was due to the fact that the Rodin tool has no capability to create proof scripts and reuse them whenever needed (such as implemented in HOL ~\cite{11} , Isabelle ~\cite{9}, PVS ~\cite{10}. Therfore, in some cases we had to manually repeat very similar or almost identical proofs.\\

\begin{small}
\begin{table}[!t]
\renewcommand{\arraystretch}{1.3}
\caption{Proof Statistics}
\label{Proof Statistics}
\centering
\begin{tabular}{|l||c||c||c|}
\hline
Model & Number of Proof  & Automatically  & Interactively \\
&Obligations&Discharged&Discharged\\
\hline
Context & 4 & 4(100\%) & 0(0\%)\\
Initial Model & 26 & 15(58\%) & 11(42\%)\\
1st Refinement & 19 & 13(68\%) & 6(32\%)\\
2nd Refinement & 95 & 34(36\%) & 61(64\%)\\
3rd Refinement & 35 & 32(91\%) & 3(9\%)\\
\hline Total & 179 & 98(54\%) & 81(46\%)\\
\hline
\end{tabular}
\end{table}
\end{small}

\section{Contribution of the paper}
The recovery algorithm that we are modeling in this paper was introduced and simulated in~\cite{5}. One part of our contribution here consists in proving  its successful termination. More importantly, we set up a formal model of arbitrary WSANs that can evolve dynamically by adding nodes and their corresponding links as well as by removing nodes and their links. Regarding the actor link recovery, we show that it is possible in three different forms that successively refine each other. Although we do not present it in this paper, these three recovery forms can be completely separated from each other in distinct Event-B events.

Specifically, we formalized the direct actor-actor recovery that relies on the global network information provided by the $closure$ construct. Moreover, we uccessfully proved termination of this formal recovery. In the next two refinements, we specified indirect actor-actor recovery via arbitrary nodes or via sensors. In these two case, we do not need the global network information to perform the recovery but rely instead on the information stored locally i.e., by the $l\_net$ relation. Since these two forms of indirect recovery are correct refinements of the direct actor-actor recovery, we can deduce that the distributed recovery is also successfully terminating. Finally the last refinement step (unfortunatelly omited due to lack of space) introduces the physical distance information into our model.

As a result our formal development acheived a complete formalization of the original algorithm presented in~\cite{5}. Our model presented in the paper demonstrates the the power and applicabability of the formal refinement approach. The original algorithm in~\cite{5} consists only of the third recovery form, with actors,  sensors, the $l\_net$ relation as well as the physical distance information detailed in~\cite{TR}. While this form is quite complex to model and prove to terminate, we have shown how to start from a more abstract version and prove the termination for it. The stepwise refinement of this initial model
 added the required complexity while keeping the desired termination property valid.

The more general message suggested by the results of this paper is that they can apply to any network with two categories of nodes, some more powerful
 than the others and coordinating with all the rest. The algorithm we have modeled is essentially a general one that can be reused as the basis for more  complex networks. We therefore aim towards having a collection of pre-proven templates that can be reused in similar situations. This aim is in agreement with creating the collection of parameterized refinement patterns for Event-B which is one of the goals of the DEPLOY project.

\section{Conclusions}
In this paper, we have formalized a distributed recovery algorithm in Event-B. The algorithm addresses the network partitioning problem in WSANs generated by actor failures. We have modeled the algorithm and the correspondent actor coordination links at three increasing levels of abstraction that refine each other. We have proved the refinement formally using the theorem prover tool Rodin~\cite{8}. The most interesting aspect put forward with our refinement modeling is the development of an actor coordination link that can be seen in three forms: a direct actor-actor link, an indirect, not further specified path, or an indirect path through sensor nodes. We have developed this link as a refinement with the precise purpose of replacing the first form with the second then third one. However, the refinement shows that all the three forms can be present in a network and thus provide various coordination alternatives for actors. In this respect, one can define coordination classes, e.g., for delegating the most security sensitive coordination to the direct actor-actor coordination links, the least real-time constrained coordination to indirect links, and the safety critical coordination to both direct actor links and indirect sensor paths between actors. This observation can prove very useful in practice.

Using the sensor infrastructure as temporary backup for actor coordination also aligns with the growing \emph{sustainability} research of using resources without depleting them. Upon detecting a direct actor-actor coordination link between two actor nodes, all sensor nodes contributing to a communication link between these actor nodes should be released of their backup task, a feature outside the scope of this paper.

Our formal WSAN model is the first attempt at formalizing WSAN algorithms in Event-B and hence the WSAN model can be much extended. For instance, non-deterministically adding and removing nodes is a useful feature for these networks as it models their dynamic scalability mechanism as well as their uncontrollable failures. However, non-deterministically adding links is just an abstraction for nodes detecting each other in wireless range and connecting via various protocols. Hence, the WSAN formal modeling space is quite generous and we intend to investigate it further, e.g., by modeling various temporal properties as well as real-time aspects and verifying various other algorithms too.

\bibliographystyle{eptcs}

\end{document}